\documentclass{article}
\usepackage{ijcai22}

\usepackage{times}
\usepackage{soul}
\usepackage{url}
\usepackage[hidelinks]{hyperref}
\usepackage[utf8]{inputenc}
\usepackage[small]{caption}
\usepackage{graphicx}
\usepackage{amsmath}
\usepackage{amsthm}
\usepackage{booktabs}
\usepackage{algorithm}
\usepackage{algorithmic}
\usepackage{amsfonts,amssymb}
\usepackage{bm}
\newcommand{\eat}[1]{}

\usepackage{multirow}
\usepackage[normalem]{ulem}
\useunder{\uline}{\ul}{}
\title{SimCGNN: Simple Contrastive Graph Neural Network for Session-based Recommendation}

\author{
Yuan Cao$^{1,2}$
\and
Xudong Zhang$^{1,2}$
\and
Fan Zhang$^{1,2}$
\and
Feifei Kou$^{1}$
\and
Josiah Poon$^{3}$
\and
Xiongnan Jin$^{4}$
\and
Yongheng Wang$^{4}$
\And
Jinpeng Chen$^{1,2, *}$
\affiliations
$^1$School of Computer Science (National Pilot Software Engineering School), Beijing University of Posts and Telecommunications, Beijing, China\\
$^2$Key Laboratory of Trustworthy Distributed Computing and Service (BUPT), Ministry of Education, Beijing, China\\
$^3$ School of Computer Science, The University of Sydney, Sydney, Australia\\
$^4$ Zhejiang Lab, Zhejiang, China
\emails
\{caoyuanboy, zhangxudong, zhang\_fan, koufeifei000, jpchen\}@bupt.edu.cn, josiah.poon@sydney.edu.au, \{xiongnan.jin, wangyh\}@zhejianglab.com
}

\begin{document}

\maketitle

\begin{abstract}
Session-based recommendation (SBR) problem, which focuses on next-item prediction for anonymous users, has received increasingly more attention from researchers. Existing graph-based SBR methods all lack the ability to differentiate between sessions with the same last item, and suffer from severe popularity bias. Inspired by nowadays emerging contrastive learning methods, this paper presents a \underline{Sim}ple \underline{C}ontrastive \underline{G}raph \underline{N}eural \underline{N}etwork for Session-based Recommendation (SimCGNN). In \textit{SimCGNN}, we first obtain normalized session embeddings on constructed session graphs. We next construct positive and negative samples of the sessions by two forward propagation and a novel negative sample selection strategy, and then calculate the constructive loss. Finally, session embeddings are used to give prediction. Extensive experiments conducted on two real-word datasets show our \textit{SimCGNN} achieves a significant improvement over state-of-the-art methods.

\end{abstract}

\section{Introduction}

With the progressive development of the contemporary internet and the explosion of information on the Internet, recommender systems have become an essential component. Sequential recommender systems consider the dynamic preference development and take both user-level information and item-level information into consideration. In certain scenarios, however, the user can be anonymous, which means that we only have access to item-level features and user-level features are not visible. Session-based recommender, which observes only the current session rather than sufficient historical user-item interaction records, has drawn attention in recent years.
\begin{figure}[tb]
    \centering
    \includegraphics[width=0.75\linewidth]{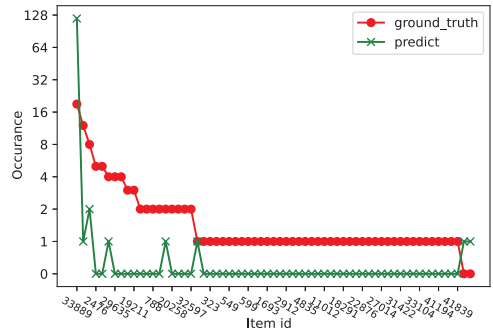}
    \caption{We trained SR-GNN on the diginetica dataset and selected a set of sessions (127 in total) with the same last-interacted item (\textit{item 33889}, the most commonly occurred last-item) from the diginetica dataset for prediction. This line graph counts the number of occurrences of different items on the ground truth and model predicted values. \eat{It is easy to see that the model predicts the next interaction item for most of the sessions as \textit{item 33889} (119/127), except for a very few extreme values (8/127). As for the ground truth, item 33889, although also the next most popular item, only accounted for 19/127. Other items that appear frequently in ground truths, i.e., \textit{item 33890}, \textit{item 32638}, are completely ignored by the SR-GNN model.}}
    \label{fig:demo}
\end{figure}

To capture the sequential relationship, Markov Chain (MC)-based sequential recommenders\cite{DBLP:conf/icdm/fossil}\cite{DBLP:conf/www/fpmc} are the first to be proposed. However, due to limited representation ability and the strong reliance upon the last interacted item in each session, their performances are limited. Recurrent neural networks (RNNs) are then introduced into session-based recommendation thanks to their natural ability for modeling sequential information. With the introduction of GRU4Rec\cite{DBLP:journals/corr/gru4rec}, RNN turned out to be the structure of choice for solving the session recommendation problem, for example, NARM\cite{DBLP:conf/cikm/narm} designs two RNNs to capture user's global and local sequential preference correspondingly. It was not until the existence of SR-GNN\cite{DBLP:conf/aaai/srgnn}, which utilized a gated graph neural network (GGNN) to extract sequential information on a session graph. Since then, works following the basic schema of SR-GNN have been proposed. TAGNN\cite{DBLP:conf/sigir/tagnn} added a target-attentive mechanism onto SR-GNN and gain promising performance. Also, as SR-GNN focuses only on a local session, methods like FGNN\cite{DBLP:journals/tois/fgnn} and GCE-GNN\cite{DBLP:conf/sigir/gce-gnn} both utilize global information in session-based recommendation but in different ways.

Although all methods mentioned above have favorable results, they maintain a drawback inherited from the MC-based approach to the present day. That is the strong dependency upon the last interacted item. Take the classical SR-GNN as an example, when assembling the final session, it simply takes linear transformation over the concatenation of the embedding vector of the last-item and the global embedding vector. And as a result, these methods may have trouble in distinguishing between sessions that have the same last interacted item. As shown in Fig.\ref{fig:demo}, we first trained an SR-GNN on the Diginetica dataset and then visualized the distribution of predicted labels and ground-truth labels of sessions that shares the same last-interacted item. This indicates a phenomenon that methods with this assembling techniques tends to predict the same items for sessions with the same last-item, despite of the underlying diversity. As we have mentioned earlier, a large branch of existing graph-based methods (\cite{DBLP:conf/sigir/tagnn,DBLP:conf/sigir/gce-gnn} for example) refers to the same approach as SR-GNN in the final assembly of the session embedding, resulting in this phenomenon will also be widespread in the current advanced graph-based SBR method. In this paper, we refer to this phenomenon as the "same last-item confusion" problem. Apart from this, Fig.\ref{fig:demo} also proved the hypothesis proposed by \cite{DBLP:journals/corr/niser} that the prediction of SR-GNN is likely to be affected by popularity bias.

To cope with the last-item, intuitively we tend to increase the discrepancy between sessions with the same last-item, which naturally fits the schema of constractive learning methods by considering sessions with the same last-item as negative samples. However, as positive samples are hard to define, we simply follows the idea of \cite{DBLP:conf/emnlp/simcse} by treating the session itself as a positive sample through dropout\cite{DBLP:journals/jmlr/dropout} techniques. To this end, we proposed a novel session-based recommender namely \underline{Sim}ple \underline{C}ontrastive \underline{G}raph \underline{N}eural \underline{N}etwork for Session-based Recommendation (\textit{SimCGNN}). \eat{which utilizes contrastive learning related techiques to increase the discrepancy between sessions with the same last-interacted item and leverage normalizations to deal with popularity bias problem.}Firstly, in order deal with the "same last-item confusion" problem, we designed a novel contrastive module to increase the discrepancy between sessions with the same last interacted item. Secondly, to eliminate the advantage of popular items in the final prediction and to predict the underlying interest of the users themselves as much as possible, we normalised both the embedding of items and sessions.

Our main contributions in this paper are listed as follows.

    1) We introduce a contrast learning approach to solve the last-item interaction confusion problem and propose a novel negative sample selection strategy.
    
    2) To alleviate the popularity bias, we proposed normalized item embeddings and session embeddings.
    
    3) Extensive experiments conducted on real-world datasets show that SimCGNN outperforms the sota methods and additional experiments also demonstrate the effectiveness of our approach for both of these two problems.

\section{Related Work}

\eat{By different network structures, we divide nowadays session-based recommenders into traditional, deep learning-based methods and neural network on graphs.}

\subsection{Traditional Methods}

A series of approaches \cite{DBLP:conf/www/MF1,DBLP:conf/nips/MF2,DBLP:journals/computer/MF3} based on matrix factorization (MF) are representatives of the traditional methods. MF-based methods consider that the user-item interaction matrix is too sparse and then decompose the matrix into two low-rank dense matrices, which correspond to users and items respectively. However, MF-based approaches do not model users' sequential interaction behavior, and therefore not suitable for making sequential recommendations.

Given that static matrix decomposition methods do not model sequence behavior, FPMC\cite{DBLP:conf/www/fpmc} models user behavior as a Markov Chain and combines Markov Chains with MF. FOSSIL\cite{DBLP:conf/icdm/fossil} improves on FPMC by introducing factorized sequential prediction with an item similarity model and higher-order Markov Chains, thus both long-term and short-term user behaviour are taken into account. However, due to expressive capability limitation of their shallow network, they do not perform well in nowadays more complex recommendation scenarios.

\subsection{Deep Learning-Based Methods}

In recent years, as deep-learning methods emerge, deep learning-based recommendation methods\cite{DBLP:conf/cikm/bert4rec,DBLP:conf/recsys/trans4rec} utilizing recurrent neural networks (RNNs) are on the rise. GRU4Rec\cite{DBLP:journals/corr/gru4rec} is a typical pioneer in utilizing RNN structure for making sequential recommendations. After that, more attempts have been made to perform sequential recommendations on RNNs. Tan et al.\cite{DBLP:conf/recsys/gru4rec+} improved the performance of RNN recommendation models by proposing several data enhancements and training tricks. Li et al.\cite{DBLP:conf/cikm/narm} proposed a neural attentive recommendation machine with an encoder-decoder structure based on RNN to capture the user's sequential preference. STAMP\cite{DBLP:conf/kdd/stamp}, which deposits the RNN structure by using simple MLPs and attention mechanisms, proved to be efficient in capturing both users' static and dynamic interests.
\eat{In addition to RNN, Caser\cite{DBLP:conf/wsdm/caser} used a convolution neural network (CNN) as the main structure for modeling sequential interaction and also obtained promising results. As Self-attention and Transformer\cite{DBLP:conf/nips/attentionisallyouneed} show their unmatched ability in the field of NLP, methods inspired by them have emerged. SAS4Rec\cite{DBLP:conf/icdm/sasrec} utilizes a self-attention mechanism as a replacement for RNN to extract internal relevance between items. Bert4Rec\cite{DBLP:conf/cikm/bert4rec} incorporates a pre-trained BERT for a sequential recommendation. It treats the users interaction sequences as text sequences and makes the final prediction. }Although deep learning-based models have more powerful representation capabilities than traditional methods, these approaches caanot still model complex item relationships, for example, non-adjacent item transitions.

\subsection{Neural Network on Graphs}

Recently, thanks to the rise of graph neural networks\cite{DBLP:journals/corr/gaT,DBLP:conf/kdd/heteroGNN,DBLP:conf/ijcai/heterolr}, many approaches\cite{DBLP:conf/ijcai/gcsan,DBLP:conf/mir/satori,DBLP:conf/sigir/lightgcn} use GNN-based network structures to solve SBR problems. SR-GNN\cite{DBLP:conf/aaai/srgnn} is one of the earliest and most representative ones, which models each session as a session graph, and applies a gated-GNN\cite{DBLP:journals/corr/ggnn} to finally get a representative embedding of the session. After the great success of SR-GNN, many variants of SR-GNN have been proposed. For example, TAGNN\cite{DBLP:conf/sigir/tagnn} proposed a target-aware attention mechanism upon SR-GNN, which adaptively activates different user interests concerning varied target items. FGNN\cite{DBLP:journals/tois/fgnn} takes both sequence order and the latent order in the session graph into consideration. Disen-GNN\cite{DBLP:journals/corr/disengnn} constructs a disentangled session graph to discover underlying session purpose. $S^2$-DHCN\cite{DBLP:conf/aaai/s2dhcn} leverages hypergraph techniques to represent each session and utilizes constractive learning techniques to perform self-supervised learning. Instead of utilizing local session graphs only, methods such as GCE-GNN\cite{DBLP:conf/sigir/gce-gnn} construct global graphs to obtain global information on the dataset.  However, as mentioned in the previous section, all of these methods have serious "same last-item confusion" problem and thus suffer from performance degradation.

\section{Methodology}

\subsection{Problem Formulation}

A session-based recommender is supposed to give recommendations to users based on the inputs of their anonymous historical interaction sequences, e.g. clicks or purchases. Given $\mathcal{V} = \left\{v_1, v_2, ..., v_m\right\}$ as the item set, an anonymous interaction session $s$ is an item sequence $s_i = [v_1^i, v_2^i, ...,v_{|s_i|}^i], s_i \in \mathcal{S}$, where $v_j^i$ is the $j$-th interacted item of the $i$-th session. Given session $s_i$, our goal is to predict $v_{|s_i|+1}^i$.

\subsection{Overview}
\begin{figure*}[tb]
    \centering
    \includegraphics[width=\linewidth]{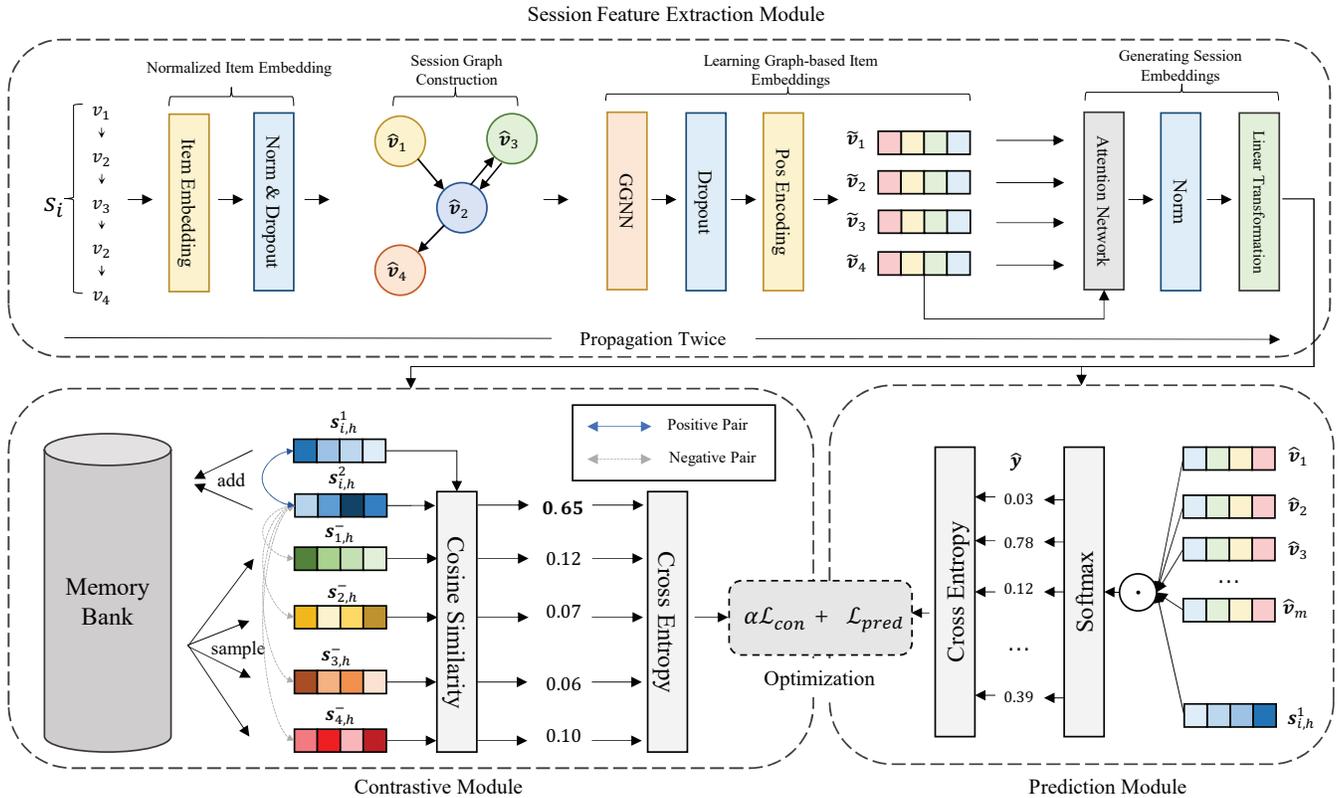}
    \caption{The overall network architecture of \emph{SimCGNN}.}
    \label{fig:overall}
\end{figure*}
The overall workflow of our proposed SimCGNN is illustrated in Fig.\ref{fig:overall}. \eat{SimCGNN consists of three main modules, i.e., Session Feature Extraction Module, Prediction Module, Contrastive Module and finally Optimization. At first, the orginal sessions are used to build session graphs, after which we leverage gated graph neural networks (GGNN) to extract graph-based item representations. Attention network and normalizations are utilized to get the final hybrid session representation. We forward propagate twice on the session feature extraction module, and as we go through the Dropout layer during propagation, we get two different session representation vectors for one session. The first representation is utilized to perform a machine factorization (MF)-like item prediction. And with sampling negative samples from memory bank, the two generated session representations are used as query and key respectively to calculate contrastive loss.}

\subsection{Session Feature Extraction Module}
\eat{Given the anonymous sessions, the first thing we need to do is to extract informative features from it. The overall feature extraction structure is similar to \cite{DBLP:conf/aaai/srgnn}, but we add fundamental modifications to 1) fitting the contrastive learning schema, 2) reducing popularity bias.}

\subsubsection{Normalized Item Embedding}

Since the one-hot vectors of items are sparse and high-dimensional and do not carry pairwise distance information, we first embed each item $v_i \in \mathcal{V}$ into a $d$-dimensional representation $\bm{v}_i \in \mathbb{R}^d$. 

As some existing methods directly utilize $\bm{v}_i$ for downstream tasks, we argue that the direct use of embedded vectors would lead to popularity bias \cite{DBLP:journals/corr/niser}. To this end, we intuitively add $L_2$ normalization to the raw feature vector. Apart from this, we also add Dropout directly on embedding layer for the downstream contrastive module. The normalized vectors $\hat{\bm{v}}_i$ are given as follows:

\begin{equation}
    \hat{\bm{v}}_i = Dropout(Norm(\bm{v}_i), p)
\end{equation}

where $p$ is the dropout probability, $Norm(\cdot)$ is the L2-normalization function.

\subsubsection{Session Graph Construction}


To explore rich transitions among items and generate accurate latent vectors of items\eat{, as shown in Fig.\ref{fig:graphcon}}, we model each session $s$ as a directed graph $\mathcal{G}_s = (\mathcal{V}_s, \mathcal{E}_s)$, where each node $ v_{s,i} \in \mathcal{V}_s$ represents an item and each edge $(v_{s, i-1}, v_{s,i}) \in \mathcal{E}_s$ represents a user interact $v_{s,i}$ after the interaction of $v_{s,i-1}$. To deal with items that occurs multiple times in sessions, we assign a normalized weight to each edge. The weight is calculated by the occurrence of the corresponding interaction ($(v_{s, i-1}, v_{s,i})$) divides the out-degree of the starting node ($v_{s, i-1}$). We use the previously obtained embedding vector $\hat{\bm{v}}_i$ as the initial state of the nodes in the constructed session graph. After the construction, we obtain the outgoing adjacency matrix $\mathbf{A}_i^{out} \in \mathbb{R}^{d\times d}$ and incoming adjacency matrix $\mathbf{A}_i^{in} \in \mathbb{R}^{d\times d}$, where $n = |V|$. By concatenating $\mathbf{A}_i^{out}$ and $\mathbf{A}_i^{in}$, we obtain the final connection matrix $\mathbf{A}_s \in \mathbb{R}^{d\times 2d}$ of the session graph.

\subsubsection{Learning Graph-based Item Embeddings}

Once the corresponding session graph has been constructed, we will then need to use a graph neural network to extract the structured information from the graph.\eat{The graph neural network, which is designed for generating node representations on graphs, is naturally suitable for session-based recommendation problem for its ability to model complex item connections.} In \textit{SimCGNN}, we leverage a gated graph neural network (GGNN) to learn node vectors in a session graph. Formally, for node $v_{s,i}$ in graph $\mathcal{G}_s$, the update function can be formulated as,
\begin{align}
    \bm{a}_{s,i}^t &= \mathbf{A}_{s,i:}\left[ \bm{v}_1^{t-1}, \bm{v}_2^{t-1}, ... , \bm{v}_{|s_i|}^{t-1} \right] ^ \top \mathbf{H} + \bm{b},\\
    \bm{z}_{s,i}^t &= \sigma\left( \mathbf{W}_z\bm{a}_{s,i}^t + \mathbf{U}_z\bm{v}_i^{t-1} \right),\\
    \bm{r}_{s,i}^t &= \sigma\left( \mathbf{W}_r\bm{a}_{s,i}^t + \mathbf{U}_r\bm{v}_i^{t-1} \right),\\
    \overline{\bm{v}}_i^t &= \mathrm{tanh}\left(\mathbf{W}_o\bm{a}_{s,i}^t + \mathbf{U}_o\left( \bm{r}_{s,i}^t \odot \bm{v}_i^{t-1} \right)\right),\\
    \bm{v}_i^t &= \left(1-\bm{z}_{s,i}^t \right) \odot \bm{v}_i^{t-1} + \bm{z}_{s,i}^t \odot \overline{\bm{v}}_i^t,
\end{align}

where $t$ is the training step, $\mathbf{A}_{s,i:}$ is the $i$-th row of matrix $\mathbf{A}_s$ corresponding to $v_{s,i}$, $\mathbf{H} \in \mathbb{R}^{d\times 2d}$ is trainable parameter, $\bm{z}_{s,i}^t$ and $\bm{r}_{s,i}^t$ are the reset and update gates respectively, $\left[ \bm{v}_1^{t-1}, \bm{v}_2^{t-1}, ... , \bm{v}_{|s_i|}^{t-1} \right]$ is the list of item vectors in $s$, $\sigma(\cdot)$ is the sigmoid function and $\odot$ is the Hadamard product operator.

In order to add the necessary noise to the resulting graph item representation vector, we also applied Dropout to the final output vector $\bm{v}_i^l$ after $l$-layers GNN. Also, since there is no location information embedded in the graph neural network, we do the usual positional embedding on the generated item vectors. The final graph-based item embeddings $\widetilde{\bm{v}_i}$ can be calculated as,

\begin{equation}
    \widetilde{\bm{v}_i} = \bm{w}_{pos(v_i)} + Dropout(\bm{v}_i, p),
\end{equation}

where $\bm{w}_t \in \mathbb{R}^d$ is the trainable positional embedding, $pos(v_i)$ is the absolute position of the item in the session., $p$ is the Dropout probability.

\subsubsection{Generating Session Embeddings}

\eat{In previous section, we obtain on-graph item representations by feeding session graphs into the graph neural networks. However, in order to make session-based predictions, we also need to obtain a representation vector for the corresponding session.} In this section, we consider constructing the session representation vector by combining long-term preference and short-term preference.

First, intuitively, we can represent the session's short-term preference by its last-interacted item $\bm{v}_{s, |s_i|}$. Thus, for session $s_i=[v_{s,1}, v_{s,2},...,v_{s,n}]$, the short-term session embedding $\bm{s}_{i,s}$ of session $s$ can be defined as $\widetilde{\bm{v}}_{n}$.

As for the long-term preference, we consider the long-term session embedding of session graph $\mathcal{G}_s$ by aggregating all node vectors. Meanwhile, as different interaction records may own different levels of priority, we utilize the attention mechanism to gain the long-term session preference $\bm{s}_{i, l}$ as follows,
\begin{align}
    \alpha_i &= \mathrm{softmax}(\bm{q}^\top \sigma(\mathbf{W}_1\widetilde{\bm{v}}_n + \mathbf{W}_2\widetilde{\bm{v}}_i + \bm{b})), \\
    \bm{s}_{i, l} &= \sum_{i=1} ^{n} \alpha_i \widetilde{\bm{v}}_i,
\end{align}
where $\bm{q}\in\mathbb{R}^{d}$ and $\mathbf{W}_1, \mathbf{W}_2 \in\mathbb{R}^{d\times d}$ are trainable weights. \eat{Notably, we also do normalization on $\alpha_i$ by utlizing softmax function.}

Finally, we get the final hybrid embedding $\bm{s}_{i,h}$ by simply linear transform over the concatenation of the short-term and the long-term session embeddings.
\begin{equation}
    \bm{s}_{i,h} = \mathbf{W}_3[\bm{s}_{i,l}; \bm{s}_{i,s}],
\end{equation}
where $\mathbf{W}_3 \in \mathbb{R}^{2d \times d}$ is trainable parameter and $[\cdot;\cdot]$ represents concatenation operation.

\subsection{Prediction Module}

After extracting session hybrid embeddings, we adopt an MF layer to predict the relevance between the given session $s_i$ and each candidate item $v_n \in V$ by multiplying the normalized session representation $Norm(\bm{s}_{i,h})$ and normalized item embedding $Norm(\bm{v}_{n})$ for the avoidance of popularity bias, which can be defined as,

\begin{equation}
    \hat{h_n} = r * Norm(\bm{s}_{i,h}) * Norm(\bm{v}_n),
\end{equation}

As $Norm(\bm{s}_{i,h}) * Norm(\bm{v}_n)$ is equal to the cosine similarity between $\bm{s}_{i,h}$ and $\bm{v}_n$, the predicted logits are restricted to $[-1, 1]$, and the softmax score is likely to get saturated at high values for the training set. To this end, we add a scaling factor $r>1$, which is useful in practice to allow for better convergence.

Then we apply a softmax function on the output logit to get the final scaled output probability vector $\hat{\bm{y}}$,
\begin{equation}
    \hat{\bm{y}} = \mathrm{softmax}(\hat{\bm{h}}),
\end{equation}

where $\hat{\bm{h}}$ denotes the recommendation scores of all candidate items $v_n \in V$.

The prediction loss $\mathcal{L}_{pred}$ is defined by calculating the cross-entropy of the prediction and ground truth,
\begin{equation}
    \mathcal{L}_{pred} = -\sum_{i=1}^{|D|} \bm{y}_i \log (\hat{\bm{y}}) + (1-\bm{y}_i)\log(1-\hat{\bm{y}}),
\end{equation}
where $\bm{y}$ denotes the one-hot encoding vector for the ground truth item.

\subsection{Contrastive Module}
As we mentioned before, simply using the hybrid session embedding $\bm{s}_{i,h}$ for item prediction inevitably leads to the same last-item confusion problem. 

To address this, the most intuitive idea is to separate session representations with the same last-item as much as possible. Naturally, we came up with the idea of using comparative learning to solve this problem. We refer to the schema of \cite{DBLP:conf/emnlp/simcse}, and since we have added the Dropout layer above, we can simply perform twice forward propagation for each session $s_i$ to produce two hybrid embeddings. For the embedding vector obtained by the first forward propagation, we denote it by $\bm{s}_{i,h}^1$ and the second by $\bm{s}_{i,h}^2$, which constructs a "positive pair". We use all embeddings of sessions with the same last item as negative samples, which can be represented by $\bm{s}^-_{j,h}, j \in 1,2,...,N$, $N$ indicates the number of negative samples. In this way, the contrastive loss $\mathcal{L}_{con}$ can be calculated as,

\begin{equation}
    \mathcal{L}_{con} = -\sum_{i=1}^{|D|} \log \frac{e^{\mathrm{sim}(\bm{s}_{i,h}^1, \bm{s}^2_{i,h})/\tau}}{\sum_{j=1}^N\left( e^{\mathrm{sim}(\bm{s}_{i,h}^1, \bm{s}^2_{i,h})/\tau} + e^{\mathrm{sim}(\bm{s}_{i,h}^1, \bm{s}^-_{j,h})/\tau} \right)},
\end{equation}

where $\tau$ is the temperature hyperparameter, $D$ is the training dataset, and we utilize cosine similarity for $\mathrm{sim}(\cdot, \cdot)$ as follows,

\begin{equation}
    \mathrm{sim}(\bm{s}_1, \bm{s}_2) = \frac{\bm{s}_1\bm{s}_2}{||\bm{s}_1|| ||\bm{s}_2||}
\end{equation}

However, unlike \cite{DBLP:conf/emnlp/simcse}, since there are not always adequate negative sessions with the same last-item in the same training batch, we cannot directly use sessions in the same training batch as negative samples. And if we forward the corresponding negative sessions each time we calculate the contrastive loss, it would result in a huge waste of computational resources. As a result, instead of exhaustively computing these representations, similar to \cite{DBLP:conf/cvpr/WuXYL18}, we maintain a memory bank $M = \{(\bm{f}^1_i, \bm{f}^2_i)\}$. During each learning iteration, the two hybrid embeddings $\bm{s}_{i,h}^1, \bm{s}_{i,h}^2$ are updated to $M$ at the corresponding session entry $\bm{s}_{i,h}^1 \longrightarrow \bm{f}^1_i, \bm{s}_{i,h}^2 \longrightarrow \bm{f}^2_i$, and negative session embeddings can be sampled from $M$, i.e., $\bm{s}_{j,1,h}^- \longleftarrow \bm{f}_j^1, \bm{s}_{j,2,h}^- \longleftarrow \bm{f}_j^2$. This allows us to compute the comparison loss without additional forward propagation, but only with the corresponding vector in the memory bank $M$.

\subsection{Model Optimization}

In the previous section, we defined the prediction loss $\mathcal{L}_{pred}$ and the contrastive loss $\mathcal{L}_{con}$, we define our 
final loss function with the integration of both of them as follows:

\begin{equation}
    \mathcal{L} = \mathcal{L}_{pred} + \beta \mathcal{L}_{con} + \lambda||\Theta||_2^2,
\end{equation}

where $\Theta$ is all trainable parameters, $\beta$ is the hyper-parameter for balancing the contrastive module, $\lambda$ is the regularization hyper-parameter.

 
\section{Experiments}

\eat{In this section, we carry out the experiments on two real-world datasets and prove the effectiveness of our proposed \textit{SimCGNN} approach outperforms. Moreover, we have designed relevant experiments to demonstrate that our approach does indeed solve for last-item confusion and popularity bias.}

\subsection{Datasets}
Following previous work, we choose two commonly used session recommendation datasets, that is, Yoochoose 1/64\footnote{http://2015.recsyschallenge.com/challege.html} and Diginetica\footnote{http://cikm2016.cs.iupui.edu/cikm-cup}. The Yoochoose dataset is from the RecSys Challenge 2015 and consists of six months of interaction sessions from an E-commercial website. We only make use of the most recent fractions 1/64 of the training sequences of Yoochoose denoted. The Diginetica dataset comes from CIKM Cup 2016, and only the transaction data is used.

For fairness consideration, we use the same preprocessing techniques as \cite{DBLP:conf/aaai/srgnn}, which filter out all sessions of length 1 and items that appears less than 5 times in both datasets. Moreover, we do data augmentation on a session to obtain sequences and corresponding labels. For example, for session $s = [v_1, v_2, ..., v_n]$, we split it in to $([v_1], v_2), ([v_1, v_2], v_3), ..., ([v_1, v_2, ..., v_{n-1}], v_n)$. The statistics of the two datasets are shown in Table \ref{table:dataset}.

\begin{table}[]
\centering
\caption{Statistic of the Datasets}
\label{table:dataset}
\begin{tabular}{@{}lll@{}}
\toprule
Dataset             & Yoochoose 1/64 & Diginetica \\ \midrule
\#click             & 557,248        & 982,961    \\
\#training sessions & 369,859        & 719,470    \\
\#test sessions     & 55,898         & 60,858     \\
\#items             & 16,766         & 43,097     \\
Average Length      & 6.16           & 5.12       \\ \bottomrule
\end{tabular}
\end{table}

\subsection{Evaluation Metrics}
According to previous works, \textbf{Recall@20} and \textbf{MRR@20} are selected to evaluate the performance of our method and baselines.

\subsection{Baselines}

To evident the effectiveness of our proposed \textit{SimCGNN}, we compare it with the following representative baselines.

\begin{itemize}
    \item \textit{POP} and \textit{SPOP} recommend the top-K popular item in training dataset and in the current predicting session respectively.
    \item \textit{Item-KNN}\cite{DBLP:conf/www/itemknn} leverages item-to-item collaborative filtering techniques, which recommends items similar to the previously interacted items by consine similarity.
    \item \textit{BPR}\cite{DBLP:conf/uai/bpr} is a classical matrix factorization (MF) methods, and is optimized by a pairwise ranking loss function.
    \item \textit{FPMC}\cite{DBLP:conf/www/fpmc} is a sequential prediction method combining the Markov chain and MF.
    \item \textit{GRU4REC}\cite{DBLP:journals/corr/gru4rec} utilizes the RNN structure to model the sequential interaction of users and leverages multiple tricks to help the RNN converge to the session recommendation problem.
    \item \textit{NARM}\cite{DBLP:conf/cikm/narm} improves GRU4REC by incorporating an attention mechanism into RNN.
    \item \textit{STAMP}\cite{DBLP:conf/kdd/stamp} replaces the RNN structures by employing attention mechanism. \eat{It captures users' short-term interests by fully relying on the self-attention of the last item.}
    \item \textit{SR-GNN}\cite{DBLP:conf/aaai/srgnn} first introduces session graph structure into a session-based recommendation. By utilizing a Gated GNN to extract on-graph item embeddings, the last item together with a weighted sum of all the session embeddings are concatenated for prediction.
    \item \textit{FGNN}\cite{DBLP:journals/tois/fgnn} formulates the next item recommendation within the session as a graph classification problem.
    \item \textit{GCE-GNN}\cite{DBLP:conf/sigir/gce-gnn} constructed a global graph using session data and modifies the model structure to introduce the global information learned from the global graph.
    \item \textit{Disen-GNN}\cite{DBLP:journals/corr/disengnn} proposes a disentangled graph neural network to capture the session purpose.
    \item \textit{$S^2$-DHCN}\cite{DBLP:conf/aaai/s2dhcn} constructs each session as a hypergraph and utilizes contractive learning methods.
\end{itemize}

\subsection{Implementation Details}

To align with the previous work, we set the hidden size $d=100$, while model parameters are initialized using  Gaussian distribution with a mean of 0 and deviation of 0.1. The mini-batch Adam optimizer is utilized to optimize model parameters. The initial learning rate is set to 1e-3 and decays by 0.1 every 3 epochs. Dropout probabilities for all dropout layers are set to 0.1. For the contrastive module, the temperature parameter $\tau$ is set to 12, and $\beta$ is set to 0.1 for Yoochoose 1/64 and 1 for Diginetica. All the hyperparameters are tuned on the validation set.

\subsection{Experiment Results}

\begin{table}[]
\centering
\caption{Recommendation performance on two datasets. The best performing method in each column is \textbf{boldfaced}, and the second best method is \underline{underlined}.}

\resizebox{0.5\textwidth}{34mm}{
\begin{tabular}{@{}ccccc@{}}
\toprule
\multirow{2}{*}{Method} & \multicolumn{2}{c}{Yoochoose 1/64} & \multicolumn{2}{c}{Diginetica}  \\
                        & Recall@20        & MRR@20          & Recall@20      & MRR@20         \\ \midrule
POP                     & 6.71             & 1.65            & 0.89           & 0.20           \\
S-POP                   & 30.44            & 18.35           & 21.06          & 13.68          \\
Item-KNN                & 51.60            & 21.81           & 35.75          & 11.57          \\
BPR                     & 31.31            & 12.08           & 5.24           & 1.98           \\
FPMC                    & 45.62            & 15.01           & 26.53          & 6.95           \\ \midrule
GRU4REC                 & 60.64            & 22.89           & 29.45          & 8.33           \\
NARM                    & 68.32            & 28.63           & 49.70          & 16.17          \\
STAMP                   & 68.74            & 29.67           & 45.64          & 14.32          \\ \midrule
SR-GNN                  & 70.57            & 30.94           & 50.73          & 17.59          \\
FGNN                    & \textbf{71.75}   & {\ul 31.71}  & 51.36          & 18.47          \\
GCE-GNN                 & 70.91            & 30.63           & \textbf{54.22} & \textbf{19.04} \\
Disen-GNN               & 71.46            & 31.36           & 53.79       & 18.99 \\
$S^2$-DHCN              & 70.39            & 29.92           & 53.66       & 18.51 \\
\midrule
\textit{SimCGNN}        & {\ul 71.61}      & \textbf{31.99}     & {\ul 54.01}    & \textbf{19.04} \\
\bottomrule
\end{tabular}}
\label{tabel:res}
\end{table}

To further demonstrate the overall performance of our proposed SimCGNN, we compare it with the selected baselines described above. The experimental results are shown in Table \ref{tabel:res}. From Table \ref{tabel:res}, we can say that our \textit{SimCGNN} achieves the best performance on all two datasets, especially in \textbf{MRR@20}, which illustrates the superior ranking capability compared with other baseline methods. 

Among traditional methods, the Item-KNN achieves the best performance, although the overall performance of all traditional methods is relatively poor. To our surprise, the simple yet effective S-POP shows better performance than those of BPR and FPMC. Notably, the S-POP takes only item popularity into consideration, which means both the Yoochoose 1/64 dataset and the Diginetica dataset are suffered from popularity bias. It's also vital to point out that, the Item-KNN utilizes pairwise item similarities only and performs better than FPMC, which is an MC-based approach with the assumption that only the last interacted items are needed to perform sequential recommendation. This phenomenon indicates that the simple MC assumption is not suitable for such complex sessions.

As for Deep learning-based methods, all DL-based methods consistently outperform traditional methods. GRU4REC and NARM are both based on RNN structure, and they together achieved decent performances. However, since NARM adds an attention mechanism to the original RNN to give different weights to items at different positions in the session, the performance of NARM has a significant improvement over GRU4REC. STAMP, which replaces RNN with attentional MLPs, shows comparative performance over NARM. At last, all DL-base methods share better performance over FPMC, which shows the importance of modeling the whole interaction sequence instead of considering merely the last click. However, both RNNs and MLPs are not suitable for capturing complex transitions among sessions. This may be the reason why they perform worse than graph-based methods.

Graph-based methods outperform all other baselines by a large margin. More specifically, GCE-GNN outperforms SR-GNN as it effectively leverages global information in different ways. FGNN also shows competitive results by rethinking item order and replacing the assembling method by a well-designed readout function. Disen-GNN gains decent performance on both datasets, showing the importance of disentangled session graphs. The $S^2-DHCN$, however, performs the worst on Yoochoose 1/64, which does not match the capacity of hypergraph neural networks. Compared with these state-of-the-art methods, our methods achieve comparative performances or even better performances without explicitly applying global information. Expanding on this, our method has optimal \textbf{MRR@20} on both datasets. The \textbf{Recall@20} of our method is higher than GCE-GNN and lower than FGNN on the Yoochoose 1/64 dataset and the opposite on the Diginetica dataset. We must also point out that the number of parameters used in our \textit{SimCGNN} is consistent with the SR-GNN. This means that we have achieved a top-level performance using the least parameters among all graph-based methods, which strongly demonstrates the superiority of our network structure. In conclusion, graph-based methods have an inherent advantage over traditional methods and DL-based methods in modeling complex sessions. \eat{Also, our \textit{SimCGNN} performs best among all graph-based methods since it to some extent solves the "same last-item confusion" problem and popularity bias problem.}

\subsection{Ablation Study}
\begin{table}[]
\centering
\caption{Performance of variants of \textit{SimCGNN}.}
\resizebox{0.5\textwidth}{18mm}{
\begin{tabular}{@{}ccccc@{}}
\toprule
\multirow{2}{*}{Method} & \multicolumn{2}{c}{Yoochoose 1/64} & \multicolumn{2}{c}{Diginetica}  \\
                        & Recall@20        & MRR@20          & Recall@20      & MRR@20         \\ \midrule
SR-GNN                  & 70.57            & 30.94           & 50.73          & 17.59          \\ \midrule
\textit{SimCGNN}        & {\ul 71.61}   & \textbf{31.99}     & \textbf{54.01} & \textbf{19.04} \\
-Contrast               & 71.31      & {\ul 31.80}  & 53.49          & {\ul 19.01}    \\
-WeakNeg                & \textbf{71.65}      & 31.16  & {\ul 53.99}          & 18.92    \\
-Norm                   & 71.01            & 31.55           & 51.77          & 17.58          \\
-PE                     & 71.39                & 30.92               & 53.80    & 18.94          \\ \bottomrule
\end{tabular}}
\label{table:abla}
\end{table}

To further validate the effectiveness of each module in our \textit{SimCGNN}, we compare our \textit{SimCGNN} with the following four variants.

\begin{itemize}
    \item \textbf{SimCGNN-Contrast}. We removed the contrastive module of \textit{SimCGNN} to prove its effectiveness.
    \item \textbf{SimCGNN-WeakNeg}. We randomly sample negative sessions instead of choosing sessions with the same last item.
    \item \textbf{SimCGNN-Norm}. We removed all normalizations to prove the effectiveness of normalized session embeddings.
    \item \textbf{SimCGNN-PE}. We removed positional embeddings to validate whether the positional information is useful for session graphs.
\end{itemize}
The results of the proposed ablation studies are shown in Table \ref{table:abla}.

\eat{\subsubsection{Influence of Contrastive Module}}
As the most essential component of our approach, we first verified the effectiveness of the proposed contrastive module. It is not difficult to see from the experimental results that \textit{SimCGNN}-Contrast performs weaker than the original \textit{SimCGNN} on both datasets. This demonstrates the effectiveness of using the contrastive related approach to enhance the representation ability of session embeddings.


In \textit{SimCGNN}-WeakNeg, we did not emphasize the importance of negative sessions with the same last-item, which greatly affected the ranking performance of the model on both datasets. In terms of recall metrics, the model was not affected too much, and there was even a marginal increase on the Yoochoose 1/64 dataset. This illustrates that our sampling method can give more favorable rankings to items that are more suitable for the target session in a relatively similar set of candidate items.

As shown in Table \ref{table:abla}, the overall performance of \textit{SimCGNN}-Norm is severely damaged. This is not only because normalization is effective in suppressing popularity bias, but also because removing normalization makes it more difficult for the contrastive module to learn the intrinsic discrepancies between sessions.

We finally verified the effectiveness of introducing positional information into the session graph. Experimental results demonstrate that the ranking performance of the model, especially on the Yoochoose 1/64 dataset, is greatly affected after the removal of the positional embedding.

\subsection{Case Study}
\subsubsection{Effect on Solving Same Last-Item Confusion}

\begin{figure}[tb]
    \centering
    \includegraphics[width=0.75\linewidth]{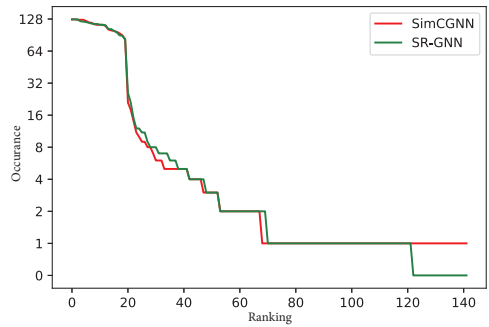}
    \caption{Predicted results of SimCGNN and SR-GNN on sessions with the same last-item (\textit{item 33889}).}
    \label{fig:demo_2}
\end{figure}

Consistent with the Introduction section, we still take sessions with the same last-item (item \textit{33889}) and utilize the trained SR-GNN and \textit{SimCGNN} to provide predictions. For each session, 20 items are recommended and counted. The relationship between the number of occurrences of an item in the prediction candidate set and its corresponding occurrence ranking is shown in Fig.\ref{fig:demo_2}. From the figure we can see that \textit{SimCGNN} predicts a lower curve of item occurrences compared to SR-GNN, which shows that SimCGNN can provide different recommendations for sessions with the same last-item. At the same time, \textit{SimCGNN} recommended a total of 142 kinds of items, 20 more than the 122 kinds of items in SR-GNN, which also demonstrates the effectiveness of our method in solving the same last-item confusion.

\subsubsection{Effect on Solving Popularity Bias}
To measure the popularity of recommended items, we calculated \textbf{Average Recommendation Popularity (ARP)} on SR-GNN and \textit{SimCGNN} to compare the difference in popularity of recommended items between the two methods. The ARP can be calculated as follows,
\begin{equation}
    ARP = \frac{1}{|S|} \sum_{s\in S}\frac{\sum_{i\in L_s} \phi(i)}{K},
\end{equation}

where $\phi(i)$ is the number of times that item $i$ appears in the training dataset, $L_s$ is the recommended item list for session $s$, and $S$ is the set of sessions in the test dataset.

From Table \ref{table:ARP}, it is easy to see that our \textit{SimCGNN} method has a huge difference in the popularity of recommended items on both datasets compared to SR-GNN. This proves that our method can indeed solve the problem of popularity bias to some extent.

\begin{table}[]
\centering
\caption{\textit{SimCGNN} versus SR-GNN in terms of Average Recommendation Popularity (ARP). Lower ARP value indicates lower popularity bias.}
\begin{tabular}{@{}ccc@{}}
\toprule
Method           & Yoochoose 1/64   & Diginetica \\ \midrule
SR-GNN           & 4128.54          & 495.25     \\
\textit{SimCGNN} & \textbf{2678.64} & \textbf{285.85}     \\ \bottomrule
\end{tabular}
\label{table:ARP}
\end{table}

\section{Conclusion}

In this paper, to address the session-based recommendation problem, we proposed a novel Simple Contrastive Graph Neural Network (\textit{SimCGNN}), which introduces a contrastive module to deal with the "same last-item confusion" problem and normalized item and session embeddings to cope with popularity bias. Experiments on two real-world datasets validate that our \textit{SimCGNN} outperforms the state-of-the-art approaches with a significant margin in terms of \textbf{Recall@20} and \textbf{MRR@20}. In future work, we aim to propose a new combination approach that can eliminate the impact of the last item to a greater extent.





\bibliographystyle{named}
\bibliography{ijcai22}

\begin{thebibliography}{}

\bibitem[\protect\citeauthoryear{Chang \bgroup \em et al.\egroup
  }{2015}]{DBLP:conf/kdd/heteroGNN}
Shiyu Chang, Wei Han, Jiliang Tang, Guo{-}Jun Qi, Charu~C. Aggarwal, and
  Thomas~S. Huang.
\newblock Heterogeneous network embedding via deep architectures.
\newblock In Longbing Cao, Chengqi Zhang, Thorsten Joachims, Geoffrey~I. Webb,
  Dragos~D. Margineantu, and Graham Williams, editors, {\em Proceedings of the
  21th {ACM} {SIGKDD} International Conference on Knowledge Discovery and Data
  Mining, Sydney, NSW, Australia, August 10-13, 2015}, pages 119--128. {ACM},
  2015.

\bibitem[\protect\citeauthoryear{Chen \bgroup \em et al.\egroup
  }{2022}]{DBLP:conf/mir/satori}
Jinpeng Chen, Yuan Cao, Fan Zhang, Pengfei Sun, and Kaimin Wei.
\newblock Sequential intention-aware recommender based on user interaction
  graph.
\newblock In Vincent Oria, Maria~Luisa Sapino, Shin'ichi Satoh, Brigitte
  Kerherv{\'{e}}, Wen{-}Huang Cheng, Ichiro Ide, and Vivek~K. Singh, editors,
  {\em {ICMR} '22: International Conference on Multimedia Retrieval, Newark,
  NJ, USA, June 27 - 30, 2022}, pages 118--126. {ACM}, 2022.

\bibitem[\protect\citeauthoryear{de Souza Pereira~Moreira \bgroup \em et
  al.\egroup }{2021}]{DBLP:conf/recsys/trans4rec}
Gabriel de~Souza Pereira~Moreira, Sara Rabhi, Jeong~Min Lee, Ronay Ak, and Even
  Oldridge.
\newblock Transformers4rec: Bridging the gap between {NLP} and sequential /
  session-based recommendation.
\newblock In Humberto Jes{\'{u}}s~Corona Pamp{\'{\i}}n, Martha~A. Larson,
  Martijn~C. Willemsen, Joseph~A. Konstan, Julian~J. McAuley, Jean
  Garcia{-}Gathright, Bouke Huurnink, and Even Oldridge, editors, {\em RecSys
  '21: Fifteenth {ACM} Conference on Recommender Systems, Amsterdam, The
  Netherlands, 27 September 2021 - 1 October 2021}, pages 143--153. {ACM},
  2021.

\bibitem[\protect\citeauthoryear{Dong \bgroup \em et al.\egroup
  }{2020}]{DBLP:conf/ijcai/heterolr}
Yuxiao Dong, Ziniu Hu, Kuansan Wang, Yizhou Sun, and Jie Tang.
\newblock Heterogeneous network representation learning.
\newblock In Christian Bessiere, editor, {\em Proceedings of the Twenty-Ninth
  International Joint Conference on Artificial Intelligence, {IJCAI} 2020},
  pages 4861--4867. ijcai.org, 2020.

\bibitem[\protect\citeauthoryear{Gao \bgroup \em et al.\egroup
  }{2021}]{DBLP:conf/emnlp/simcse}
Tianyu Gao, Xingcheng Yao, and Danqi Chen.
\newblock Simcse: Simple contrastive learning of sentence embeddings.
\newblock In Marie{-}Francine Moens, Xuanjing Huang, Lucia Specia, and
  Scott~Wen{-}tau Yih, editors, {\em Proceedings of the 2021 Conference on
  Empirical Methods in Natural Language Processing, {EMNLP} 2021, Virtual Event
  / Punta Cana, Dominican Republic, 7-11 November, 2021}, pages 6894--6910.
  Association for Computational Linguistics, 2021.

\bibitem[\protect\citeauthoryear{Gupta \bgroup \em et al.\egroup
  }{2019}]{DBLP:journals/corr/niser}
Priyanka Gupta, Diksha Garg, Pankaj Malhotra, Lovekesh Vig, and Gautam Shroff.
\newblock {NISER:} normalized item and session representations with graph
  neural networks.
\newblock {\em CoRR}, abs/1909.04276, 2019.

\bibitem[\protect\citeauthoryear{He and McAuley}{2016}]{DBLP:conf/icdm/fossil}
Ruining He and Julian~J. McAuley.
\newblock Fusing similarity models with markov chains for sparse sequential
  recommendation.
\newblock In Francesco Bonchi, Josep Domingo{-}Ferrer, Ricardo Baeza{-}Yates,
  Zhi{-}Hua Zhou, and Xindong Wu, editors, {\em {IEEE} 16th International
  Conference on Data Mining, {ICDM} 2016, December 12-15, 2016, Barcelona,
  Spain}, pages 191--200. {IEEE} Computer Society, 2016.

\bibitem[\protect\citeauthoryear{He \bgroup \em et al.\egroup
  }{2020}]{DBLP:conf/sigir/lightgcn}
Xiangnan He, Kuan Deng, Xiang Wang, Yan Li, Yong{-}Dong Zhang, and Meng Wang.
\newblock Lightgcn: Simplifying and powering graph convolution network for
  recommendation.
\newblock In Jimmy~X. Huang, Yi~Chang, Xueqi Cheng, Jaap Kamps, Vanessa
  Murdock, Ji{-}Rong Wen, and Yiqun Liu, editors, {\em Proceedings of the 43rd
  International {ACM} {SIGIR} conference on research and development in
  Information Retrieval, {SIGIR} 2020, Virtual Event, China, July 25-30, 2020},
  pages 639--648. {ACM}, 2020.

\bibitem[\protect\citeauthoryear{Hidasi \bgroup \em et al.\egroup
  }{2016}]{DBLP:journals/corr/gru4rec}
Bal{\'{a}}zs Hidasi, Alexandros Karatzoglou, Linas Baltrunas, and Domonkos
  Tikk.
\newblock Session-based recommendations with recurrent neural networks.
\newblock In Yoshua Bengio and Yann LeCun, editors, {\em 4th International
  Conference on Learning Representations, {ICLR} 2016, San Juan, Puerto Rico,
  May 2-4, 2016, Conference Track Proceedings}, 2016.

\bibitem[\protect\citeauthoryear{Koren \bgroup \em et al.\egroup
  }{2009}]{DBLP:journals/computer/MF3}
Yehuda Koren, Robert~M. Bell, and Chris Volinsky.
\newblock Matrix factorization techniques for recommender systems.
\newblock {\em Computer}, 42(8):30--37, 2009.

\bibitem[\protect\citeauthoryear{Li \bgroup \em et al.\egroup
  }{2016}]{DBLP:journals/corr/ggnn}
Yujia Li, Daniel Tarlow, Marc Brockschmidt, and Richard~S. Zemel.
\newblock Gated graph sequence neural networks.
\newblock In Yoshua Bengio and Yann LeCun, editors, {\em 4th International
  Conference on Learning Representations, {ICLR} 2016, San Juan, Puerto Rico,
  May 2-4, 2016, Conference Track Proceedings}, 2016.

\bibitem[\protect\citeauthoryear{Li \bgroup \em et al.\egroup
  }{2017}]{DBLP:conf/cikm/narm}
Jing Li, Pengjie Ren, Zhumin Chen, Zhaochun Ren, Tao Lian, and Jun Ma.
\newblock Neural attentive session-based recommendation.
\newblock In Ee{-}Peng Lim, Marianne Winslett, Mark Sanderson, Ada~Wai{-}Chee
  Fu, Jimeng Sun, J.~Shane Culpepper, Eric Lo, Joyce~C. Ho, Debora Donato,
  Rakesh Agrawal, Yu~Zheng, Carlos Castillo, Aixin Sun, Vincent~S. Tseng, and
  Chenliang Li, editors, {\em Proceedings of the 2017 {ACM} on Conference on
  Information and Knowledge Management, {CIKM} 2017, Singapore, November 06 -
  10, 2017}, pages 1419--1428. {ACM}, 2017.

\bibitem[\protect\citeauthoryear{Li \bgroup \em et al.\egroup
  }{2022}]{DBLP:journals/corr/disengnn}
Ansong Li, Zhiyong Cheng, Fan Liu, Zan Gao, Weili Guan, and Yuxin Peng.
\newblock Disentangled graph neural networks for session-based recommendation.
\newblock {\em CoRR}, abs/2201.03482, 2022.

\bibitem[\protect\citeauthoryear{Liu \bgroup \em et al.\egroup
  }{2018}]{DBLP:conf/kdd/stamp}
Qiao Liu, Yifu Zeng, Refuoe Mokhosi, and Haibin Zhang.
\newblock {STAMP:} short-term attention/memory priority model for session-based
  recommendation.
\newblock In Yike Guo and Faisal Farooq, editors, {\em Proceedings of the 24th
  {ACM} {SIGKDD} International Conference on Knowledge Discovery {\&} Data
  Mining, {KDD} 2018, London, UK, August 19-23, 2018}, pages 1831--1839. {ACM},
  2018.

\bibitem[\protect\citeauthoryear{Qiu \bgroup \em et al.\egroup
  }{2019}]{DBLP:journals/tois/fgnn}
Ruihong Qiu, Jingjing Li, Zi~Huang, and Hongzhi Yin.
\newblock Rethinking the item order in session-based recommendation with graph
  neural networks.
\newblock pages 579--588, 2019.

\bibitem[\protect\citeauthoryear{Rendle \bgroup \em et al.\egroup
  }{2009}]{DBLP:conf/uai/bpr}
Steffen Rendle, Christoph Freudenthaler, Zeno Gantner, and Lars
  Schmidt{-}Thieme.
\newblock {BPR:} bayesian personalized ranking from implicit feedback.
\newblock In Jeff~A. Bilmes and Andrew~Y. Ng, editors, {\em {UAI} 2009,
  Proceedings of the Twenty-Fifth Conference on Uncertainty in Artificial
  Intelligence, Montreal, QC, Canada, June 18-21, 2009}, pages 452--461. {AUAI}
  Press, 2009.

\bibitem[\protect\citeauthoryear{Rendle \bgroup \em et al.\egroup
  }{2010}]{DBLP:conf/www/fpmc}
Steffen Rendle, Christoph Freudenthaler, and Lars Schmidt{-}Thieme.
\newblock Factorizing personalized markov chains for next-basket
  recommendation.
\newblock In Michael Rappa, Paul Jones, Juliana Freire, and Soumen Chakrabarti,
  editors, {\em Proceedings of the 19th International Conference on World Wide
  Web, {WWW} 2010, Raleigh, North Carolina, USA, April 26-30, 2010}, pages
  811--820. {ACM}, 2010.

\bibitem[\protect\citeauthoryear{Salakhutdinov and
  Mnih}{2007}]{DBLP:conf/nips/MF2}
Ruslan Salakhutdinov and Andriy Mnih.
\newblock Probabilistic matrix factorization.
\newblock In John~C. Platt, Daphne Koller, Yoram Singer, and Sam~T. Roweis,
  editors, {\em Advances in Neural Information Processing Systems 20,
  Proceedings of the Twenty-First Annual Conference on Neural Information
  Processing Systems, Vancouver, British Columbia, Canada, December 3-6, 2007},
  pages 1257--1264. Curran Associates, Inc., 2007.

\bibitem[\protect\citeauthoryear{Sarwar \bgroup \em et al.\egroup
  }{2001a}]{DBLP:conf/www/MF1}
Badrul~Munir Sarwar, George Karypis, Joseph~A. Konstan, and John Riedl.
\newblock Item-based collaborative filtering recommendation algorithms.
\newblock In Vincent~Y. Shen, Nobuo Saito, Michael~R. Lyu, and Mary~Ellen
  Zurko, editors, {\em Proceedings of the Tenth International World Wide Web
  Conference, {WWW} 10, Hong Kong, China, May 1-5, 2001}, pages 285--295.
  {ACM}, 2001.

\bibitem[\protect\citeauthoryear{Sarwar \bgroup \em et al.\egroup
  }{2001b}]{DBLP:conf/www/itemknn}
Badrul~Munir Sarwar, George Karypis, Joseph~A. Konstan, and John Riedl.
\newblock Item-based collaborative filtering recommendation algorithms.
\newblock In Vincent~Y. Shen, Nobuo Saito, Michael~R. Lyu, and Mary~Ellen
  Zurko, editors, {\em Proceedings of the Tenth International World Wide Web
  Conference, {WWW} 10, Hong Kong, China, May 1-5, 2001}, pages 285--295.
  {ACM}, 2001.

\bibitem[\protect\citeauthoryear{Srivastava \bgroup \em et al.\egroup
  }{2014}]{DBLP:journals/jmlr/dropout}
Nitish Srivastava, Geoffrey~E. Hinton, Alex Krizhevsky, Ilya Sutskever, and
  Ruslan Salakhutdinov.
\newblock Dropout: a simple way to prevent neural networks from overfitting.
\newblock {\em J. Mach. Learn. Res.}, 15(1):1929--1958, 2014.

\bibitem[\protect\citeauthoryear{Sun \bgroup \em et al.\egroup
  }{2019}]{DBLP:conf/cikm/bert4rec}
Fei Sun, Jun Liu, Jian Wu, Changhua Pei, Xiao Lin, Wenwu Ou, and Peng Jiang.
\newblock Bert4rec: Sequential recommendation with bidirectional encoder
  representations from transformer.
\newblock In Wenwu Zhu, Dacheng Tao, Xueqi Cheng, Peng Cui, Elke~A.
  Rundensteiner, David Carmel, Qi~He, and Jeffrey~Xu Yu, editors, {\em
  Proceedings of the 28th {ACM} International Conference on Information and
  Knowledge Management, {CIKM} 2019, Beijing, China, November 3-7, 2019}, pages
  1441--1450. {ACM}, 2019.

\bibitem[\protect\citeauthoryear{Tan \bgroup \em et al.\egroup
  }{2016}]{DBLP:conf/recsys/gru4rec+}
Yong~Kiam Tan, Xinxing Xu, and Yong Liu.
\newblock Improved recurrent neural networks for session-based recommendations.
\newblock In Alexandros Karatzoglou, Bal{\'{a}}zs Hidasi, Domonkos Tikk,
  Oren~Sar Shalom, Haggai Roitman, Bracha Shapira, and Lior Rokach, editors,
  {\em Proceedings of the 1st Workshop on Deep Learning for Recommender
  Systems, DLRS@RecSys 2016, Boston, MA, USA, September 15, 2016}, pages
  17--22. {ACM}, 2016.

\bibitem[\protect\citeauthoryear{Velickovic \bgroup \em et al.\egroup
  }{2017}]{DBLP:journals/corr/gaT}
Petar Velickovic, Guillem Cucurull, Arantxa Casanova, Adriana Romero, Pietro
  Li{\`{o}}, and Yoshua Bengio.
\newblock Graph attention networks.
\newblock {\em CoRR}, abs/1710.10903, 2017.

\bibitem[\protect\citeauthoryear{Wang \bgroup \em et al.\egroup
  }{2020}]{DBLP:conf/sigir/gce-gnn}
Ziyang Wang, Wei Wei, Gao Cong, Xiao{-}Li Li, Xianling Mao, and Minghui Qiu.
\newblock Global context enhanced graph neural networks for session-based
  recommendation.
\newblock In Jimmy~X. Huang, Yi~Chang, Xueqi Cheng, Jaap Kamps, Vanessa
  Murdock, Ji{-}Rong Wen, and Yiqun Liu, editors, {\em Proceedings of the 43rd
  International {ACM} {SIGIR} conference on research and development in
  Information Retrieval, {SIGIR} 2020, Virtual Event, China, July 25-30, 2020},
  pages 169--178. {ACM}, 2020.

\bibitem[\protect\citeauthoryear{Wu \bgroup \em et al.\egroup
  }{2018}]{DBLP:conf/cvpr/WuXYL18}
Zhirong Wu, Yuanjun Xiong, Stella~X. Yu, and Dahua Lin.
\newblock Unsupervised feature learning via non-parametric instance
  discrimination.
\newblock In {\em 2018 {IEEE} Conference on Computer Vision and Pattern
  Recognition, {CVPR} 2018, Salt Lake City, UT, USA, June 18-22, 2018}, pages
  3733--3742. Computer Vision Foundation / {IEEE} Computer Society, 2018.

\bibitem[\protect\citeauthoryear{Wu \bgroup \em et al.\egroup
  }{2019}]{DBLP:conf/aaai/srgnn}
Shu Wu, Yuyuan Tang, Yanqiao Zhu, Liang Wang, Xing Xie, and Tieniu Tan.
\newblock Session-based recommendation with graph neural networks.
\newblock In {\em The Thirty-Third {AAAI} Conference on Artificial
  Intelligence, {AAAI} 2019, The Thirty-First Innovative Applications of
  Artificial Intelligence Conference, {IAAI} 2019, The Ninth {AAAI} Symposium
  on Educational Advances in Artificial Intelligence, {EAAI} 2019, Honolulu,
  Hawaii, USA, January 27 - February 1, 2019}, pages 346--353. {AAAI} Press,
  2019.

\bibitem[\protect\citeauthoryear{Xia \bgroup \em et al.\egroup
  }{2021}]{DBLP:conf/aaai/s2dhcn}
Xin Xia, Hongzhi Yin, Junliang Yu, Qinyong Wang, Lizhen Cui, and Xiangliang
  Zhang.
\newblock Self-supervised hypergraph convolutional networks for session-based
  recommendation.
\newblock In {\em Thirty-Fifth {AAAI} Conference on Artificial Intelligence,
  {AAAI} 2021, Thirty-Third Conference on Innovative Applications of Artificial
  Intelligence, {IAAI} 2021, The Eleventh Symposium on Educational Advances in
  Artificial Intelligence, {EAAI} 2021, Virtual Event, February 2-9, 2021},
  pages 4503--4511. {AAAI} Press, 2021.

\bibitem[\protect\citeauthoryear{Xu \bgroup \em et al.\egroup
  }{2019}]{DBLP:conf/ijcai/gcsan}
Chengfeng Xu, Pengpeng Zhao, Yanchi Liu, Victor~S. Sheng, Jiajie Xu, Fuzhen
  Zhuang, Junhua Fang, and Xiaofang Zhou.
\newblock Graph contextualized self-attention network for session-based
  recommendation.
\newblock In Sarit Kraus, editor, {\em Proceedings of the Twenty-Eighth
  International Joint Conference on Artificial Intelligence, {IJCAI} 2019,
  Macao, China, August 10-16, 2019}, pages 3940--3946. ijcai.org, 2019.

\bibitem[\protect\citeauthoryear{Yu \bgroup \em et al.\egroup
  }{2020}]{DBLP:conf/sigir/tagnn}
Feng Yu, Yanqiao Zhu, Qiang Liu, Shu Wu, Liang Wang, and Tieniu Tan.
\newblock {TAGNN:} target attentive graph neural networks for session-based
  recommendation.
\newblock In Jimmy~X. Huang, Yi~Chang, Xueqi Cheng, Jaap Kamps, Vanessa
  Murdock, Ji{-}Rong Wen, and Yiqun Liu, editors, {\em Proceedings of the 43rd
  International {ACM} {SIGIR} conference on research and development in
  Information Retrieval, {SIGIR} 2020, Virtual Event, China, July 25-30, 2020},
  pages 1921--1924. {ACM}, 2020.

\end{thebibliography}

\end{document}